\documentclass[conference,9pt]{IEEEtran}
\IEEEoverridecommandlockouts
\usepackage{cite}
\usepackage{amsmath,amssymb,amsfonts,bm}
\usepackage{paralist}
\usepackage{float}
\usepackage{algorithmic}
\usepackage{graphicx}
\usepackage{textcomp}
\usepackage{xcolor}
\def\BibTeX{{\rm B\kern-.05em{\sc i\kern-.025em b}\kern-.08em
    T\kern-.1667em\lower.7ex\hbox{E}\kern-.125emX}}

\usepackage{array}
\usepackage[ruled, linesnumbered,noend]{algorithm2e}
\usepackage{comment}

\usepackage[skip=0pt,font=smaller]{subcaption}
\usepackage{array}
\usepackage{multirow}
\usepackage[bookmarks=true,breaklinks=true,letterpaper=true,colorlinks,citecolor=blue,linkcolor=blue,urlcolor=blue]{hyperref}

\newcolumntype{A}{>{\centering\arraybackslash}m{2.2cm}}
\newcolumntype{B}{>{\centering\arraybackslash}m{1.7cm}}
\newcolumntype{H}{>{\centering\arraybackslash}m{0.55cm}}
\newcolumntype{J}{>{\centering\arraybackslash}m{1.35cm}}
\newcolumntype{K}{>{\centering\arraybackslash}m{1.4cm}}

\newcolumntype{E}{>{\centering\arraybackslash}m{2.2cm}}
\newcolumntype{F}{>{\centering\arraybackslash}m{1.7cm}}

\newcommand*{\myfont}{\fontfamily{pcr}\selectfont}
\DeclareTextFontCommand{\myemph}{\myfont}

\newfloat{Algorithm}{htbp}{loa}
\begin{document}

\newcommand{\mntJoinSpeedup}{$4.65\times$}
\newcommand{\mntRegSpeedup}{$7.46\times$}
\newcommand{\pimdbSpeedup}{$1.83\times$}
\newcommand{\twoxbSpeedup}{$3.39\times$}
\newcommand{\twoxbSpeedupMntJoin}{$1.37\times$}
\newcommand{\pimdbEnergyGeomean}{$4.31\times$}
\newcommand{\pimdbEnergyL}{$2.98\times$}
\newcommand{\pimdbEnergyH}{$6.38\times$}
\newcommand{\pimdbPowerGeomean}{$2.92\times$}
\newcommand{\pimdbPowerH}{$23.27\times$}
\newcommand{\pimdbEnduGeomean}{$3.21\times$}
\newcommand{\pimdbEnduL}{$2.98\times$}
\newcommand{\pimdbEnduH}{$17.15\times$}

\title{Enabling Relational Database Analytical Processing in Bulk-Bitwise Processing-In-Memory
\vspace{-10pt}
\thanks{This work was supported by the European Research Council through the European Union's Horizon 2020 Research and Innovation Programme under Grant 757259 and through the European Union's Horizon Europe Research and Innovation Programme under Grant 101069336.}
}

\author{\IEEEauthorblockN{Ben Perach \qquad Ronny Ronen \qquad Shahar Kvatinsky}
\IEEEauthorblockA{\textit{Viterbi Faculty of Electrical \& Computer Engineering} \\
Technion -- Israel Institute of Technology, 
Haifa, Israel \\
benperach@technion.ac.il \quad ronny.ronen@ef.technion.ac.il \quad shahar@ee.technion.ac.il\vspace{-18pt}}
}

\maketitle

\begin{abstract}

Bulk-bitwise processing-in-memory (PIM), an emerging computational paradigm utilizing memory arrays as computational units, has been shown to benefit database applications. This paper demonstrates how GROUP-BY and JOIN, database operations not supported by previous works, can be performed efficiently in bulk-bitwise PIM for relational database analytical processing. We extend the gem5 simulator and evaluated our hardware modifications on the Star Schema Benchmark. We show that compared to previous works, our modifications improve (on average) execution time by \pimdbSpeedup, energy by \pimdbEnergyGeomean, and the system's lifetime by \pimdbEnduGeomean. We also achieved a speedup of \mntJoinSpeedup~over MonetDB, a modern state-of-the-art in-memory database.

\end{abstract}

\section{Introduction}
\label{sec:intro}
Processing-in-memory (PIM) is an emerging computing paradigm that mitigates the latency and energy associated with data movements by computing where the data reside. In this paper, we focus on a specific PIM technique called \textit{bulk-bitwise PIM}~\cite{PIMDB,AMBIT,RACER,SIMDRAM,CONCEPT,Pinatubo}, in which the memory arrays also act as bit-vector processing units operating on their stored data.
Previous works on bulk-bitwise PIM showed that database applications, specifically relational database online analytical processing (\textit{OLAP})~\cite{Kimball2013}, can be substantially accelerated by bulk-bitwise PIM~\cite{CONCEPT,PIMDB,AMBIT,SIMDRAM,Pinatubo}. OLAP database queries process many records, often summarizing information from multiple requested records subgroups. 
The database operations accelerated by previous works, however, are limited to filtering and aggregation operations. Those works did not support additional database operations such as GROUP-BY and JOIN needed to perform entire OLAP queries. JOIN operations combine several database relations (tables) and allow their combined information to be queried. GROUP-BY operations divide the records into subgroups and then summarize each subgroup. 

This paper presents bulk-bitwise PIM techniques that support GROUP-BY and most JOIN operations for relational database OLAP. To the best of our knowledge, this is the first work that supports such operations with bulk-bitwise PIM.
Supporting JOIN requires heavy data movement. As bulk-bitwise PIM support in data movement is limited, JOIN is supported by storing pre-joined relations in the PIM memory. Pre-joining, as denormalization~\cite{Shin2006} or as materialized view~\cite{Chirkova2012}, is a known method to accelerate query execution. Pre-joining, however, adds maintenance complexity and storage overheads. We argue that pre-joined relations are suitable for OLAP on bulk-bitwise PIM, and we show how bulk-bitwise PIM can mitigate the drawbacks of pre-join.
To support GROUP-BY, we adopt an in-cloud processing GROUP-BY technique~\cite{PushdownDB} to bulk-bitwise PIM, where the work is divided between the host processor and PIM. To efficiently adapt this GROUP-BY technique to bulk-bitwise PIM, we add a circuit to the PIM memory, accelerating PIM aggregation and improving memory cell lifetime.

We implement our proposed methods on a gem5~\cite{gem5} full-system simulation (including an operating system) and measure their performance using the Star Schema Benchmark (SSB)~\cite{SSB}, a relational database analytic processing benchmark. We are unaware of any other work that has designed and evaluated a complete database benchmark on a bulk-bitwise PIM system. We compare the query execution time to MonetDB~\cite{MonetDB}, a modern in-memory database system, and achieve a geometric mean (geo-mean) speedup of \mntRegSpeedup~and \mntJoinSpeedup~over the standard and pre-joined versions of the SSB benchmark, respectively.

In summary, this paper makes the following contributions:
\begin{compactitem}
    \item We show how bulk-bitwise PIM can mitigate the drawbacks of pre-joined relations.
    \item We adapt a GROUP-BY algorithm for bulk-bitwise PIM.
    \item We add an aggregation circuit to the memory arrays' peripherals to accelerate aggregation operations, improving previous work with an average speedup of \pimdbSpeedup, improving energy by \pimdbEnergyGeomean, and improving the systems's lifetime by \pimdbEnduGeomean.
    \item We evaluate our solutions using a gem5 full-system simulation of an RRAM-based PIM system and the SSB benchmark, showing runtime improvements of \mntJoinSpeedup~and \mntRegSpeedup~over a modern in-memory database with and without pre-joining relations, respectively.
\end{compactitem}

\vspace{-6pt}
\section{Background}

\subsection{Relational Databases and Analytical Processing}

A relational database is a data model that organizes data in relations (tables)~\cite{Kimball2013}. Each relation in the database holds multiple records, viewed as the rows of the relations. Each relation has several attributes, viewed as the relations columns, where each record has a value for each attribute. A relation has one attribute or a combination of attributes designated as its \textit{key}, uniquely identifying its records.

Database queries are questions about the database records, \textit{i.e.}, returning records, or a function on records that fulfill a particular condition on their attributes.
Analytical processing, used in applications such as business decision support processes, has dedicated database structures and is characterized by specific types of queries~\cite{SSB,Kimball2013}. Such queries usually have the form \textit{select-from-where-group by}, which means that queries search for records satisfying a certain condition (\textit{where}) on one or more relations (\textit{from}), and return an aggregation (\textit{e.g.}, sum, average, max) on some attribute for these records (\textit{select}). Frequently, an aggregation is required per subgroup of the selected records, classified according to some attributes, which is referred to as GROUP-BY. 

A JOIN operation between relations is required when a query involves more than a single relation. The JOIN operation connects records of the different relations according to a condition on their attributes, allowing to check attributes from several relations together. A JOIN operation can comprise over $90\%$ of execution time in analytical processing~\cite{Dreseler2020}, but the partition into relations is kept to maintain flexibility in execution, avoid data duplication, and simplify database maintenance~\cite{Shin2006}.
\subsection{Bulk-Bitwise PIM}
\begin{figure}[!t]
\centering
\begin{minipage}[c]{0.05\columnwidth}
\begin{subfigure}[c]{\textwidth}
\caption{}\label{subfig:bitwise_logic}
\end{subfigure}
\end{minipage}%
\begin{minipage}[c]{0.42\columnwidth}
\includegraphics[width=\textwidth]{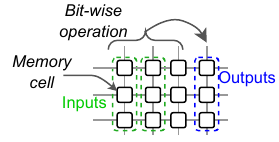}
\end{minipage}%
\begin{minipage}[c]{0.05\columnwidth}
\begin{subfigure}[c]{\textwidth}
\caption{}\label{subfig:system}
\end{subfigure}
\end{minipage}%
\begin{minipage}[c]{0.48\columnwidth}
\includegraphics[width=\textwidth]{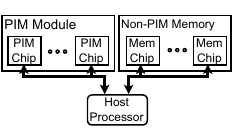}
\end{minipage}%
\vspace{-3 pt}
\caption{(a) A $3\times4$ memory crossbar array performing a bitwise column logic operation (\textit{e.g.}, NOR). The operation inputs and output are the two left column cell values (green) and the right column cells (blue). Bitwise row operations can be performed similarly. (b) A bulk-bitwise PIM module connected as a memory rank to a host.}
\label{fig:bb_pim}
\end{figure}

Bulk-bitwise PIM uses the memory arrays and their peripherals (\textit{e.g.}, decoders, sense amplifiers, and voltage drivers) as processing elements. Processing is done within the memory array, accessing input and output data directly from and to the memory cells, eliminating data movement out of the memory array. 
The basic operations supported by these memory crossbars are logic operations (\textit{e.g.}, NOR~\cite{PIMDB,RACER}).
Because of the regular structure of memory arrays, these logic operations can be performed concurrently on numerous cells with the memory array. Furthermore, many memory arrays can operate concurrently, resulting in a wide logic operation, \textit{i.e.}, bulk-bitwise operations.
An example of bulk-bitwise operation using a memory crossbar array is shown in Fig.~\ref{subfig:bitwise_logic}.
More complex operations (\textit{e.g.}, addition, multiplication) can be constructed using sequences of the basic logic operations.
DRAM~\cite{SIMDRAM,AMBIT} and emerging nonvolatile memory technologies~\cite{RACER,PIMDB,CONCEPT,Pinatubo} have been suggested to implement such bulk-bitwise PIM.  

\setlength{\textfloatsep}{0pt}

Bulk-bitwise PIM memory can be used as the main memory of a host processor~\cite{PIMDB,SIMDRAM,AMBIT}. The PIM module is constructed and serves as a memory rank with PIM-enabled memory chips in addition to standard non-PIM (\textit{e.g.}, DRAM) memory ranks (Fig.~\ref{subfig:system}). In this case, the host can read and write to/from the PIM memory using standard loads and stores. To perform a PIM computation, the host sends a memory command, named \textit{PIM request}, to the PIM module. PIM requests are sent with an address and data, similar to store instructions, detailing the computation, operands, and result location. Virtual memory is supported by restricting PIM requests to use and modify data only within a single memory page~\cite{PIMDB}. The address of the PIM request specifies this page. When issuing a PIM request, user-level programs send PIM requests with a virtual address. The virtual address is translated into a physical address using the standard translation methods and forwarded to the relevant memory location.

To enable pages to operate independently in memory, each page has a dedicated controller, named \textit{PIM controller}, on each memory chip. When a PIM request arrives at a memory chip, the PIM controller for the targeted page manages the required basic logic operation sequence to all crossbars of that page. To maintain high parallelism, huge pages (\textit{e.g.}, 2MB or 1GB) are used, operating concurrently with the same operation on all crossbars belonging to that page.  

When mapping databases to bulk-bitwise PIM memory, each relation has its pages where the relation's records are stored. Each record is set as a single crossbar row, where the attributes of all the records are aligned on crossbar columns~\cite{PIMDB,CONCEPT,AMBIT,SIMDRAM}. To filter records according to some condition, the condition is implemented by column-wise PIM operations on the same attributes on all the relation's records. The filter result is a single bit per record, indicating whether the record satisfied the filter condition. Hence, reading the filter result requires reading a single bit per record instead of the filtered attributes.
Aggregation is done by concurrently aggregating the same attribute in all the relation's crossbars. This concurrent aggregation is followed by reading the aggregated values from each crossbar and combining them at the host. Consequently, for aggregation, only a single value is read from each crossbar rather than the entire relevant attribute per record.
Due to the reduction in read operations, the substantial reduction in data movement is the main benefit of bulk-bitwise PIM and can reach $99\%$ of the reads without bulk-bitwise PIM~\cite{PIMDB}.

\vspace{-2pt}
\section{Supporting Pre-Joined Relations}
\label{sec:join}
\vspace{-2pt}
JOIN operations require matching records from two or more relations, frequently in a many-to-many or one-to-many manner. Since the matched records from different relations cannot be assumed to reside on the same PIM crossbar, data must be moved to perform a JOIN operation. Furthermore, the match (or matches) for a record depends on the record data, requiring data-dependent movement. Although data movement within the memory is possible for bulk-bitwise PIM~\cite{RACER}, it has not been shown to support virtual memory; moreover, it is not data dependent. Therefore, it cannot trivially support JOIN operations.
On the other hand, operations performed locally within a crossbar, \textit{i.e.,} operations on a single relation, will benefit from the high parallelism of bulk-bitwise PIM~\cite{PIMDB}. Consequently, we propose to keep pre-joined relations in the PIM module, enabling full queries to be performed on a single relation. 

Pre-joined relations can appear as a result of denormalization~\cite{Shin2006} or materialized views~\cite{Chirkova2012}, both of which are known methods to accelerate query execution by compromising on other aspects. These aspects are:
(1) Limited flexibility -- pre-joining relations is only sometimes helpful since there are many possible JOIN operations, and the selected pre-joined relation may not be the one required by a query.
(2) Additional storage -- the JOIN output can be larger than the sum of the input relations.
(3) Complicated maintenance -- as JOIN operations often duplicate a single datum to multiple locations, operations such as UPDATE become more complicated. 

\begin{figure}[!t]
\centering
\includegraphics[width=0.95\columnwidth]{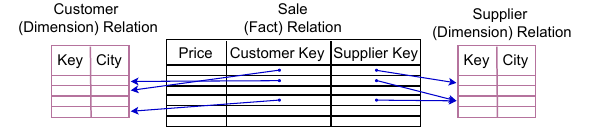}%
\vspace{-7 pt}
\caption{A star schema example for storing sales information. The database contains a single fact relation storing sales data (a record per sale) and two dimension relations storing data about customers and suppliers.}
\vspace{-3 pt}
\label{fig:star_schema}
\end{figure}

Regarding the limited flexibility aspect, we note that for the \textit{star schema}, an OLAP database structure common in business support processes~\cite{SSB,Kimball2013}, there are only a few frequently used JOIN options. A star schema contains a single large central relation and multiple smaller relations, called \textit{fact} and \textit{dimension} relations, respectively. Fig.~\ref{fig:star_schema} shows an example of a star schema.
A record in the fact relation represents a single event with its quantitative details (\textit{e.g.}, a purchase with its price). Some of the fact relation attributes are keys of the dimension relations (\textit{i.e.}, foreign keys), connecting events to their additional information (\textit{e.g.}, customer information). Queries often require a JOIN operation between a dimension relation and the fact relation using an equality condition on the dimension keys~\cite{Kimball2013}, \textit{i.e.}, an equi-JOIN on the dimension keys. Hence, most queries in a star schema for OLAP will benefit from keeping a pre-joined relation.
The dimension relations will be pre-joined to the fact relation by equi-JOIN on dimension keys, and the flexibility will not be impeded. For example, this kind of JOIN satisfies all queries of the Star Schema Benchmark (SSB)~\cite{SSB}.

Regarding the additional storage aspect, since the JOIN operation is on the dimensions' keys, and keys are unique, the JOIN connects a fact relation record with a single dimension relation record.
Hence, the fact relation's records are not duplicated, and the JOIN's output relation has the same number of records as the fact relation. However, dimension information is added to each record from the fact relation (\textit{e.g.}, a customer's information is attached to all of his purchases), increasing the record size. For bulk-bitwise PIM, relations are stored in dedicated pages, usually underutilizing the crossbar row for each record~\cite{PIMDB}. This unused row memory can be exploited if more information for each relation record is stored there.
Hence, if bulk-bitwise PIM is used for the fact relation, storing the pre-joined relation of the fact and dimension relations can use the unused memory, as it has the same number of records as the fact relation. This results in no additional memory requirements. 

Generally, the resulting record of the pre-joined relation might be larger than a single crossbar row. In that case, the pre-joined relation can be vertically partitioned~\cite{PIMDB}, storing the relation's attributes on multiple aligned pages. Such vertical partitioning, however, will add a memory overhead and hurt performance, as intermediate results will have to be transferred between the partitions. This partition should, therefore, locate the commonly used attributes together in a single crossbar, preventing intermediate result transfers in the common case. For the SSB benchmark, however, the pre-joined relation record does not exceed row size, not requiring such partitioning. Section~\ref{subsec:methodology}, nevertheless, does evaluate this case as well.

\begin{algorithm}[t]
\small
\SetAlgoNoLine
\DontPrintSemicolon
\SetKwInOut{Input}{Input}\SetKwInOut{Output}{Output}
\Input{$v_{n-1}...v_0$ - in-memory attribute bits to update \\$c_{n-1}...c_0$ - update value bits as immediate\\
$s$ - in-memory select bit\\}
\KwResult{For all $i$: $v_i\leftarrow c_i$ if $s=1$, else $v_i$ is unchanged.}
 \BlankLine
 \ForEach{$i\in[0,..,n-1]$}{
  \eIf{$c_i = 1$}{$InMemory(v_i\leftarrow v_i\  OR\  s)$
   }{$InMemory(v_i\leftarrow v_i\  AND\  NOT(s))$
  }
 }
\caption{\small PIM controller algorithm for MUX between an in-memory and immediate values, using an in-memory select.
}
\label{algo:mux}
\end{algorithm}

Pre-joined relations, in general, inherently require complex maintenance~\cite{Shin2006,Chirkova2012}. Specifically, UPDATE operations become slower due to data duplication. With bulk-bitwise PIM, an UPDATE operation can be performed using PIM by filtering the relations records according to the to-be-replaced attribute value. The filter result is then used to overwrite the attribute of only the desired records, \textit{i.e.}, the filter result is used as the \textit{select} bit for a PIM-implemented multiplexer (MUX). The PIM MUX algorithm, inspired by~\cite{PIMDB}, is described in Alg.~\ref{algo:mux}. This UPDATE operation requires only PIM operations and no read operation, eliminating data movement almost entirely.  

\vspace{-3pt}
\section{Supporting GROUP-BY}
\label{sec:groupby}
\vspace{-2pt}
To support GROUP-BY operations, we adopt an algorithm designed for in-cloud processing~\cite{PushdownDB}. In-cloud processing and bulk-bitwise PIM~\cite{PIMDB} support the same database primitives: filtering and aggregation. Bulk-bitwise PIM, however, differs from in-cloud processing in its characteristics (\textit{e.g.}, processing latency, data retrieval latency), resulting in different behavior and parameters.

\textbf{GROUP-BY Technique:} 
After filtering the required records for a query, each subgroup for the GROUP-BY can be aggregated using two options.
The first option aggregates each subgroup separately using PIM operations. Each subgroup is further filtered and then aggregated, both using PIM operations. We name this option \textit{pim-gb}. The pim-gb operates on the entire relation, making its latency independent of the number of records in each subgroup but dependent on the relation's number of records and the number of subgroups. Although PIM aggregation has low latency, there can be many subgroups to aggregate,
leading to high latency for pim-gb. 
The second option uses the host to read the filter result and the records that pass the filter. Each record is first read, and, according to its attribute values, is assigned to a subgroup and aggregated at the host. This option handles many subgroups concurrently and does not require PIM operations in addition to the filtering specified by the query. We name this option \textit{host-gb}. The host-gb's latency mainly depends on the number of required memory reads. The number of these reads depends on the relation's number of records, the total number of subgroups' records, and the size of attributes to be read from each record.

In this work, the GROUP-BY technique exploits the fact that pim-gb and host-gb depend on different parameters. Pim-gb depends on the number of subgroups and is independent of the number of records in the subgroups, while host-gb works the other way around. 
Additionally, the GROUP-BY technique relies on the fact that database data is not uniformly distributed~\cite{Rabl2013} and the GROUP-BY subgroups have non-uniform sizes. Thus, subgroups are divided such that a few large subgroups are aggregated by pim-gb, leaving the many small remaining subgroups (which might be empty) to be handled by host-gb.
The division of subgroups between pim-gb and host-gb depends on the data distribution and the specific query requirements. To decide how to perform this division, the host samples a small fraction of the records selected by the query, and using this sample, it estimates the size of each subgroup. According to an empirical model (described later in this section), the host decides which subgroups to aggregate by PIM and which subgroups to aggregate at the host. This technique was suggested in~\cite{PushdownDB} for in-cloud processing and we adapt it here for bulk-bitwise PIM. 

\begin{figure}[!t]
\centering
\includegraphics[width=0.80\columnwidth]{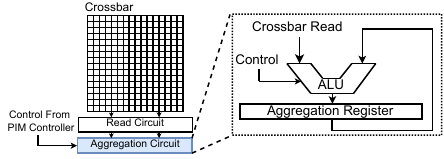}%
\vspace{-7 pt}
\caption{Circuit added to the memory crossbars (blue). The circuit aggregates the data read from the crossbar. The ALU supports the operations of SUM, MIN, and MAX.}
\vspace{-3 pt}
\label{fig:agg_circuit}
\end{figure}

\textbf{Accelerating PIM Aggregation:}
Previous work showed that bulk-bitwise PIM aggregation operations are expensive in terms of execution time, power, and cell endurance (for emerging nonvolatile memory technologies)~\cite{PIMDB}. This cost is due to the high number of basic operations required to perform aggregation. Hence, to enable efficient GROUP-BY execution, we add an arithmetic circuit to the periphery of each memory crossbar, supporting the required aggregation operations (see~\cite{PIMDB}), as shown in Fig.~\ref{fig:agg_circuit}. This approach diverges from and complements a pure bulk-bitwise PIM architecture to mitigate the weak points of aggregation in bulk-bitwise PIM.  

The arithmetic circuit receives data read from the memory crossbar. The aggregated values, specified by the PIM request, are read one by one serially, and aggregated in the arithmetic circuit. The circuit is designed with standard CMOS logic, performing only the required logic for aggregation~\cite{PIMDB}: SUM, MIN, and MAX, and controlled by the PIM controller. Since crossbar reads have a fixed bit length~\cite{Xu2015} (16 bits in our evaluation), supporting aggregation of larger word lengths requires the ALU to support the shifting and masking of its operands. The final aggregation result is then written from the arithmetic circuit to the crossbar, using a modified write control logic~\cite{Pinatubo}. The location where to write the result is specified by the aggregation PIM request sent by the host. The host can then access the final aggregation result using standard memory reads.

\textbf{Empirical Modeling:}
As stated above, the GROUP-BY technique requires deciding which subgroups to assign to pim-gb and which to host-gb. A latency model for each option is needed to make a quantitative decision. To obtain such a model, we performed latency measurements (on the system described in Section~\ref{subsec:methodology}, using synthetic databases) and fit the results of each option into a mathematical expression (as described below).
This fitting process can be automated by the database management software~\cite{Dageville2004}. In all the expressions below, $M$ is the number of 2MB pages required to store the relation.

\begin{figure}[t]
\vspace{7pt}
\centering
\begin{minipage}[c]{0.05\columnwidth}
\begin{subfigure}[c]{\textwidth}
\caption{}\label{subfig:agg_host}
\end{subfigure}
\end{minipage}%
\begin{minipage}[c]{0.85\columnwidth}
\includegraphics[width=\textwidth,trim=0 0pt 0 8pt,clip]{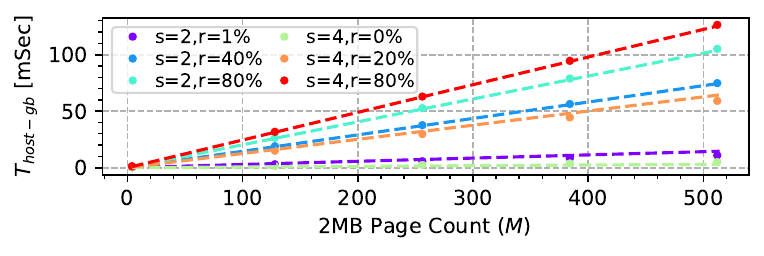}
\end{minipage}\vspace{-5pt}
\begin{minipage}[c]{0.05\columnwidth}
\begin{subfigure}[c]{\textwidth}
\caption{}\label{subfig:agg_host_slope}
\end{subfigure}
\end{minipage}%
\begin{minipage}[c]{0.42\columnwidth}
\includegraphics[width=\textwidth,trim=0 0pt 0 4pt,clip]{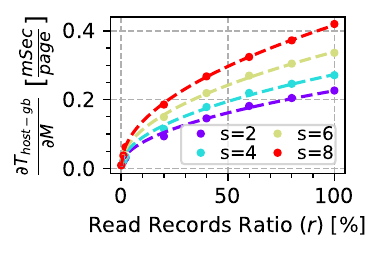}
\end{minipage}%
\begin{minipage}[c]{0.05\columnwidth}
\begin{subfigure}[c]{\textwidth}
\caption{}\label{subfig:agg_pim}
\end{subfigure}
\end{minipage}%
\begin{minipage}[c]{0.42\columnwidth}
\includegraphics[width=\textwidth,trim=0 0pt 0 4pt,clip]{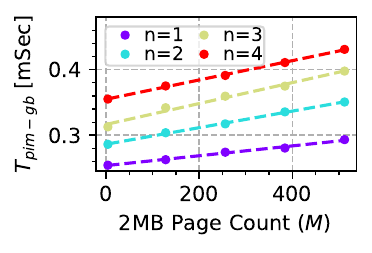}
\end{minipage}%
\vspace{-9pt}
\caption{Empirical latency modeling, showing both empirical measurements (dots) and fit (dashed lines). For brevity, only subsets of measurements are shown. (a) $T_{host{\text -}gb}$ vs. page count ($M$) for different reads per record ($s$) and read record ratio ($r$). (b) $\frac{\partial T_{host{\text -}gb}}{\partial M}$ vs. $r$ for different $s$. (c) $T_{pim{\text -}gb}$ for a single subgroup vs. page count for different aggregation attribute reads ($n$).} 
\label{fig:agg_host}
\end{figure}

The host-gb latency, $T_{host{\text -}gb}$, includes the filtering of relevant subgroups using PIM, reading the filter result (bit-vector), and reading the selected records. The filter PIM operation latency is dominated by the filter result reads~\cite{PIMDB}, whose latency depends only on the relation size. When reading the selected records, the latency depends on the number of selected records and the required reads per record. We mark with $r$ the ratio of the selected records to the total records in the relation, and with $s$ the required number of reads per record (composed of the subgroup identifiers and aggregated attributes). Hence, host-gb depends on three parameters: relation size ($M$), the ratio of records to read ($r$), and the number of reads per record ($s$). Fig.~\ref{subfig:agg_host} shows that $T_{host{\text -}gb}$, for specific $r$ and $s$ values, is linear in $M$. The slope $\frac{\partial T_{host{\text -}gb}}{\partial M}$ as a function of $r$ for a given $s$ is shown in Fig.~\ref{subfig:agg_host_slope}. For a given $s$, the slope exhibits a relation of the form $a\sqrt{r}+b$, where $a$ and $b$ are constants. Since reads from crossbars have a fixed length (16 bits in our evaluation), $s$ can have a few discrete values. Hence, we express $a$ and $b$ as lookup tables for values of $s$, receiving the following expression:
\vspace{-2pt}
\begin{equation}
    T_{host{\text -}gb}(M,s,r) = M\cdot\left( a(s)\cdot\sqrt{r}+b(s) \right).
    \vspace{-4pt}
\end{equation}

The pim-gb latency for a single subgroup ($T_{pim{\text -}gb}$) depends on the relation size ($M$) and the number of reads to retrieve the aggregated attribute from a single crossbar (marked by $n$); it is independent of the number of aggregated records. 
We measure aggregation latency on varying relation and operand sizes. The results are shown in Fig.~\ref{subfig:agg_pim}. The aggregation latency is linear in the relation size, with coefficients depending on $n$. As with $s$ above, $n$ can have a few discrete values. Hence, the coefficients $\frac{\partial T_{pim{\text -}gb}}{\partial M}$ and $T_{pim{\text -}gb,0}$ (the free coefficient) are expressed as lookup tables for values of $n$. The resulting latency fit for a single subgroup PIM aggregate is:
\vspace{-4pt}
\begin{equation}
    T_{pim{\text -}gb}(M,n) = M\cdot \frac{\partial T_{pim{\text -}gb}}{\partial M}(n) + T_{pim{\text -}gb,0}(n).
    \vspace{-4pt}
\end{equation}

To perform the GROUP-BY operation, we decide how many subgroups are assigned to pim-gb. We mark this number as $k$. By definition, these are the $k$ largest subgroups. The ratio of remaining records for host-gb to total records depends on $k$ and the data distribution, and is marked as $r(k)$. The function $r(\cdot)$ is estimated by the record sampling mentioned previously. The total GROUP-BY latency, $T_{gb}$, is, therefore:
\vspace{-4pt}
\begin{equation}
\label{eq:groupby_latency}
\begin{aligned}
    T&_{gb}(M,n,s,k_{MAX},k,r(\cdot)) =\\
    k&\cdot T_{pim{\text -}gb}(M,n)\: + \:(1-\delta_{k,k_{MAX}})\cdot T_{host{\text -}gb}\left( M,s,r(k)\right),
\end{aligned}
\vspace{-2pt}
\end{equation}
where $k_{MAX}$ is the total number of subgroups and $\delta_{i,j}$ is the Kronecker delta. The $(1-\delta_{k,k_{MAX}})$ term indicates that if all subgroups are aggregated using pim-gb, then host-gb is not performed. Using (\ref{eq:groupby_latency}), retrieving $M$, $n$, $s$, and $k_{MAX}$, from the query and database definitions, and estimating $r(\cdot)$, we find the optimal number of PIM aggregated subgroups, $k$, to achieve the lowest $T_{gb}$.

\vspace{-4pt}
\section{Evaluation}
\vspace{-6pt}
\subsection{Methodology}
\label{subsec:methodology}

\textbf{Simulation:} To evaluate our proposed methods, we developed a gem5 simulation~\cite{gem5} based on the system from~\cite{PIMDB,PIMConsistency} and avilable at~\cite{simulation}
, having an RRAM bulk-bitwise PIM, running in a full-system mode (running a Linux kernel), and including the gem5's ruby cache system. We take the coherency solution and scope consistency model from~\cite{PIMConsistency}. 
The system parameters are listed in Table~\ref{table:arch_details}. 

\begin{table}[t]
\setlength{\tabcolsep}{2pt}
\small
\vspace{7pt}
\centering
\begin{tabular}{|A|B||E|F|} 
\hline
\multicolumn{4}{|c|}{\rule{0pt}{1.6ex}Single RRAM PIM Module}\\
\hline
\rule{0pt}{1.5ex} Total Capacity & 32GB & Huge pages size  & 2MB  \\\hline
\rule{0pt}{1.5ex}   Memory ranks & 1 & PIM Chips & 8  \\\hline
\rule{0pt}{1.5ex}   Crossbar rows  & 1024  & Crossbar columns  & 512   \\\hline
\rule{0pt}{1.5ex}   Crossbar read  & 16 bit  & Bulk-bitwise logic cycle  &  30 ns~\cite{CONCEPT}  \\\hline
\rule{0pt}{1.5ex}   Crossbar read/write energy  & \mbox{0.84\textbackslash6.9} pJ/bit~\cite{CONCEPT}  & Bulk-bitwise logic energy  &  \mbox{81.6 fJ/bit} \cite{Talati2016} \\\hline
\rule{0pt}{1.5ex}  Single agg. circuit power   & 25.4 uW  & Single PIM controller power  &  126 uW~\cite{PIMDB} \\\hline

\multicolumn{4}{|c|}{\rule{0pt}{1.6ex}Evaluation System}\\
\hline 
\rule{0pt}{1.5ex}Processor cores& \mbox{6 cores, X86,} \mbox{OoO, 3.6GHz}  & Main memory  & \mbox{32GB DRAM}, \mbox{DDR4-2400}  \\\hline
\rule{0pt}{1.5ex}   L1 cache & \mbox{Private, 16KB,} \mbox{64B block}, 4-way  & L2 cache & \mbox{Shared, 2MB,} \mbox{64B block}, 16-way \\\hline
\rule{0pt}{1.5ex} Coherence protocol  & MESI  & RRAM PIM modules & 1 \\\hline

\end{tabular}
\vspace{-2pt}
\caption{Architecture and system configuration}
\label{table:arch_details}
\vspace{-14pt}
\end{table}

\begin{figure}[t]
\centering
\includegraphics[width=0.69\columnwidth,trim=0 15pt 0 0pt,clip]{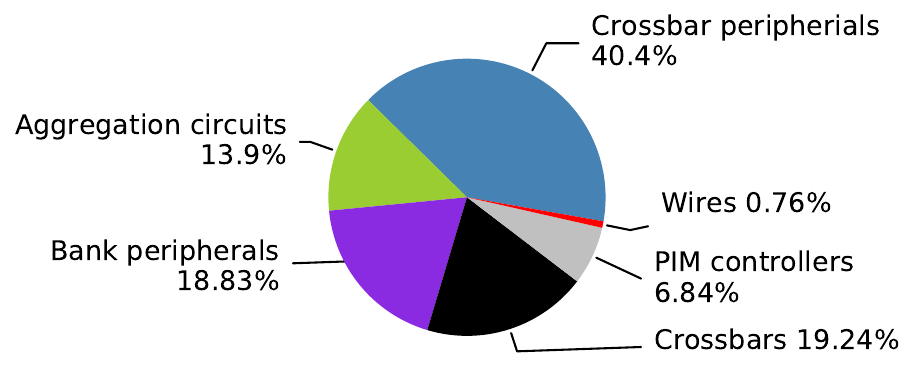} %
\caption{PIM chip area breakdown} 
\label{fig:area}
\vspace{-0pt}
\end{figure}

\begin{figure*}[t]
\vspace{4pt}
\centering
\includegraphics[width=0.87\textwidth,trim=0 0pt 0 7pt,clip]{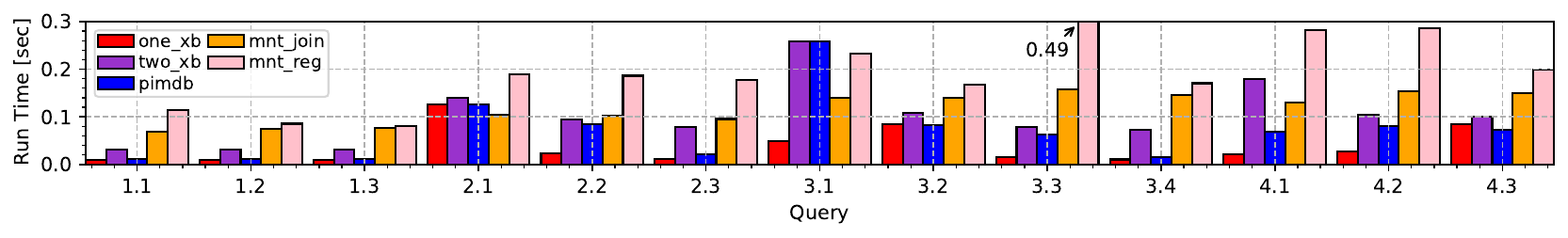} 
\vspace{-10pt}
\caption{Execution latency for the SSB benchmark queries.}
\vspace{-16pt}
\label{fig:latency}
\end{figure*}

\textbf{Benchmark}: We ran the SSB benchmark~\cite{SSB} with a scale factor of ten ($SF=10$) to evaluate the performance of our proposed methods. SSB is an OLAP benchmark containing 13 queries, divided into four query groups. We adopt the compiling procedure from~\cite{PIMDB}. Using an offline in-house compiler, the SSB SQL queries are compiled into a C++ code, which is then compiled with gcc. The query execution divides the relation records into four equal groups (in page granularity), and each group is assigned to a single thread. For a GROUP-BY operation, the record sampling and estimation described in Section~\ref{sec:groupby} are done once and shared among the four threads. The sampling is performed over a single 2MB page, \textit{i.e.,} 32K records.

The relations of the SSB benchmark are stored as a single pre-joined relation, the result of an equi-JOIN between the fact and dimensions relations on the dimension keys. We populate the relation according to~\cite{Rabl2013} with non-uniform data. When required, we change the parameters of the queries to retain similar query \textit{selectivity} (the ratio of filtered records out of the total records) as in the original uniform data~\cite{SSB}. The pre-joined relation contains all the attributes of the original relations, except the \myemph{NAME} and \myemph{ADDRESS} attributes of the \myemph{CUSTOMER} and \myemph{SUPPLIER} dimension relations. These attributes are long texts that are not used by the SSB queries. By not including these attributes, we enable a single record of the pre-joined relation to fit in a single crossbar row. Thus, the vertical partitioning described in Section~\ref{sec:join} is not required.

\begin{table}[t]
\setlength{\tabcolsep}{2pt}
\small
\centering
\begin{tabular}{|@{\hspace{0pt}}c|c|c|c|c|c|c|} 
\hline
\multirow{2}{*}{~\textbf{Q}} & \multirow{2}{*}{\textbf{Selectivity}} & \multirow{2}{*}{\parbox{1.4cm}{\centering\textbf{Total\\subgroups}}} & \multirow{2}{*}{\parbox{1.36cm}{\centering\textbf{Subgroups in sample}}} & \multicolumn{3}{c|}{\textbf{PIM agg. subgroups}} \\\cline{5-7}
&&&&\textbf{one-xb} & \textbf{two-xb} & \textbf{pimdb}  \\
\hline
\rule{0pt}{1.2ex} 1.1 & 2.3e-2  & 1     & -     & 1     & 1 & 1 \\\hline
\rule{0pt}{1.2ex} 1.2 & 6.6e-4  & 1     & -     & 1     & 1 & 1 \\\hline
\rule{0pt}{1.2ex} 1.3 & 8.4e-5  & 1     & -     & 1     & 1 & 1 \\\hline
\rule{0pt}{1.2ex} 2.1 & 1.2e-2  & 280   & 121   & 4     & 0 & 0 \\\hline
\rule{0pt}{1.2ex} 2.2 & 1.6e-3  & 56    & 33    & 56    & 0 & 0  \\\hline 
\rule{0pt}{1.2ex} 2.3 & 2e-4    & 7     & 4     & 7     & 0 & 7  \\\hline 
\rule{0pt}{1.2ex} 3.1 & 3.4e-2  & 150   & 150   & 150   & 0 & 0  \\\hline 
\rule{0pt}{1.2ex} 3.2 & 1.3e-3  & 600   & 27    & 27    & 0 & 0  \\\hline 
\rule{0pt}{1.2ex} 3.3 & 4.7e-5  & 24    & 2     & 24    & 0 & 0  \\\hline
\rule{0pt}{1.2ex} 3.4 & 6.6e-7  & 4     & 0     & 4     & 0 & 4  \\\hline
\rule{0pt}{1.2ex} 4.1 & 2e-2    & 35    & 35    & 35    & 0 & 35  \\\hline
\rule{0pt}{1.2ex} 4.2 & 2.3e-3  & 50    & 29    & 50    & 0 & 0  \\\hline
\rule{0pt}{1.2ex} 4.3 & 9.1e-5  & 800   & 3     & 3     & 0 & 0  \\\hline
\end{tabular}
\vspace{-5pt}
\caption{Query summary: \textbf{Selectivity} (the ratio of filtered records out of the total records), the total number of potential subgroups according to query and database details \textbf{(total subgroups}), subgroups found in the sampling for the GROUP-BY estimation (\textbf{subgroups in sample}), and the number of \textbf{PIM aggregated subgroups}. Q1.1--3 do not require a GROUP-BY and perform a single aggregation using PIM.}
\label{table:groupby_subgroups}
\vspace{-2pt}
\end{table}

We ran our solution in two versions. The first version, named \textit{one-xb}, holds a record of the pre-joined relation in a single crossbar row. The second version, named \textit{two-xb}, implements the vertical partitioning described in Section~\ref{sec:join}, splitting records across two crossbars, and evaluates cases where records are too large to fit in a single crossbar. All attributes of the fact relation were placed in a single crossbar; the attributes of the dimension relations were placed in a second crossbar. The vertical partitioning changes the latency of the PIM aggregation due to the required additional transfer through the host of results between the partitions~\cite{PIMDB}. For all GROUP-BY operations in SSB, the subgroup identifier attributes were from the dimension relations, and the aggregated attributes were from the fact relation, making two-xb the worst-case partitioning. To perform aggregation in PIM, the filter results for \textit{each} subgroup are transferred between pages prior to the PIM aggregation, resulting in a substantial overhead due to worst-case partitioning. If prior knowledge of common subgroup identifiers is available, the most common ones can be placed on the same crossbar with the attributes from the fact relation, reducing the required data movement. From this perspective, one-xb evaluates the best case for such a partition. We repeated the pim-gb and host-gb empirical modeling described in Section~\ref{sec:groupby} for the two-xb version.

\textbf{Area and Power:} To evaluate the area and power of the added aggregation circuit from Section~\ref{sec:groupby}, we designed the circuit using Verilog, synthesized it, and determined its area and power using Synopsys Design Compiler and Cadence Innovus with TSMC CMOS 28nm technology.
For the PIM module chip area, NVSim~\cite{NVSim} was modified to include the PIM controllers according to~\cite{PIMDB} and our aggregation circuit per crossbar. The PIM module consists of eight chips, and each occupying $346 mm^2$. Fig.~\ref{fig:area} shows the PIM module chip area breakdown, with the aggregation circuit consuming $13.9\%$ of the chip area. The area overhead of the aggregation circuit is relatively high since it is added to each crossbar.  
To estimate the PIM module power and energy on query execution, power and energy parameters of the different parts of the PIM module are taken from~\cite{PIMDB} (summarized in Table~\ref{table:arch_details}) and summed by the gem5 simulation.

\begin{figure}[t]
\centering
\includegraphics[width=0.87\columnwidth,trim=0 0pt 0 7pt,clip]{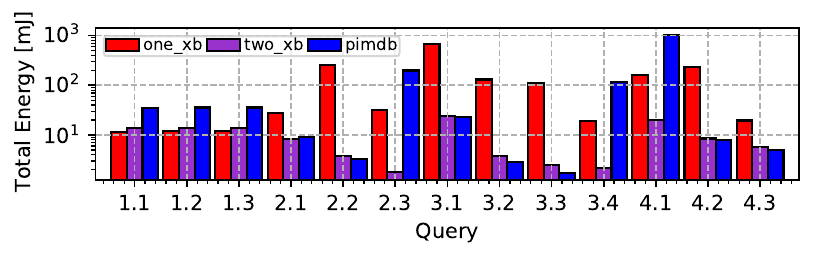}
\vspace{-10pt}
\caption{PIM memory energy for the SSB benchmark.} 
\label{fig:energy}
\vspace{-12pt}
\end{figure}

\begin{figure}[t]
\centering
\includegraphics[width=0.87\columnwidth,trim=0 0pt 0 7pt,clip]{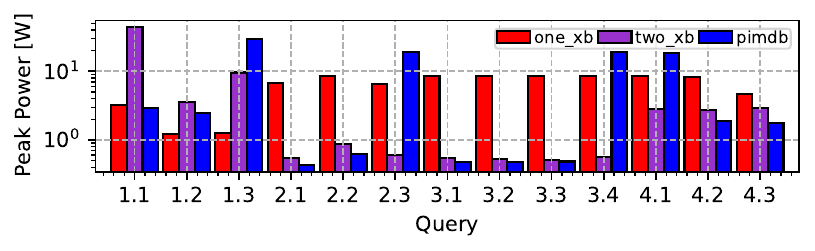}
\vspace{-10pt}
\caption{Peak power for a single PIM chip for the SSB benchmark.} 
\label{fig:power}
\vspace{-2pt}
\end{figure}

\textbf{Comparisons:} Our solution is compared against MonetDB, a modern in-memory database management system for OLAP~\cite{MonetDB}, running on a server with two Intel Xeon processors (each has 16 cores at 2.1GHz), a total of 256GB DDR4-2400 memory, and running Ubuntu. We ran two versions of MonetDB. The first version, named \textit{mnt-reg}, had the original SSB relation schema. A second version, named \textit{mnt-join}, was run with the pre-joined relation as in one-xb. The reported latency for MonetDB only includes execution time without SQL parsing and optimization latencies.

We also compared our aggregation circuit solution to PIMDB~\cite{PIMDB}, on which our system is based. We extend PIMDB to use the pre-joined relation and GROUP-BY as in one-xb. PIMDB differs only in its PIM aggregation, performing it purely with bulk-bitwise logic, while our solution uses the aggregation circuit from Section~\ref{sec:groupby}. PIMDB's PIM aggregation is empirically modeled as in Section~\ref{sec:groupby}. All other aspects of PIMDB and one-xb are identical. 
\vspace{-4pt}
\subsection{Results}
\vspace{-2pt}
\textbf{Execution Latency:} Fig.~\ref{fig:latency} shows the execution latencies for the SSB queries. The query execution summary is listed in Table~\ref{table:groupby_subgroups}. \textit{One-xb} achieves the best execution latency, having a geo-mean speedup of \mntRegSpeedup~and \mntJoinSpeedup~over mnt-reg and mnt-join, respectively.
If the pre-joined relation is vertically partitioned across two crossbars (\textit{two-xb}), there is a geo-mean slowdown of \twoxbSpeedup~compared to one-xb, which is, however, still \twoxbSpeedupMntJoin~faster than mnt-join.
PIMDB has a \pimdbSpeedup~slowdown compared to one-xb, showing the latency improvement achieved by our aggregation circuit. The improvement in aggregation latency enables the GROUP-BY technique to assign more subgroups for PIM aggregation, as shown in Table~\ref{table:groupby_subgroups}.

For the GROUP-BY queries with the highest selectivity, most notably Q2.1, Q3.1, and Q4.1, the PIM solutions exhibit low speedup and even a slowdown compared to MonetDB. Since a read from the PIM memory spans several crossbars from a page~\cite{PIMDB} and a record is in a single crossbar on a page, reading a single record brings many records from the memory to the host, 32 records in our system. This read amplification cancels out the read reduction achieved by the PIM filtering. With the low to no read reduction for PIM, the read reduction techniques of MonetDB (\textit{e.g.}, filtering, indexing) and the stronger system used for MonteDB gives it the advantage in such cases. Note that if there are sufficiently few subgroups, a pure pim-gb can be performed and achieve speedup even with high selectivity, as with one-xb in Q3.1 and Q4.1.

\textbf{Energy and Power:} The resulting energy used by the PIM module and the peak power drawn by a PIM chip are shown in Figs.~\ref{fig:energy} and~\ref{fig:power}, respectively. All queries require less than 1J for the PIM module and less than 44W peak power per PIM chip. When PIMDB uses PIM aggregation (Q1.1--1.3, Q2.3, Q3.4, and Q4.1), it consumes \pimdbEnergyGeomean~more energy in geo-mean than one-xb, and its peak power is \pimdbPowerGeomean~higher than one-xb. However, the situation is reversed in the other queries since one-xb uses PIM aggregation and PIMDB does not. As bulk-bitwise PIM performs wide operations, spanning many crossbar rows concurrently, it requires more energy in less time than read operations at the PIM module. Hence, by not performing PIM aggregation, PIMDB trades-off energy for latency. For the \textit{two-xb}, the PIM and read operations on more crossbars increase the peak power at Q1.1--1.3. In the other queries, the pure host-gb keeps its energy and peak power low.

\begin{figure}[t]
\vspace{4pt}
\centering
\includegraphics[width=0.87\columnwidth,trim=0 0pt 0 7pt,clip]{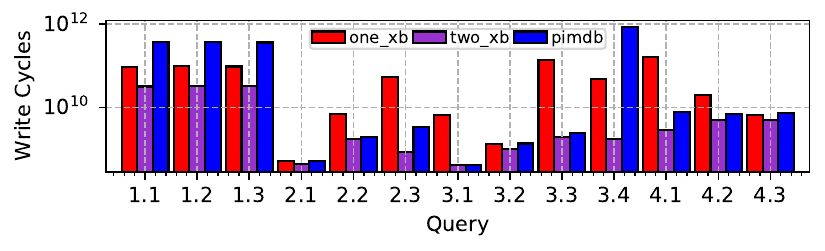}
\vspace{-10pt}
\caption{Required cell endurance for the SSB benchmark queries, assuming each query is performed back-to-back for ten years.} 
\label{fig:endurance}
\vspace{-2pt}
\end{figure}

\textbf{Endurance:} Fig.~\ref{fig:endurance} shows the required endurance for a single memory cell when running each query back-to-back ($100\%$ duty cycle) for ten years. The numbers in the figure assume that wear leveling techniques are performed, and the operations are uniformly distributed across the cells of each row~\cite{PIMDB}.
For each query, the cell usage is the maximum number of operations a single crossbar row experiences, divided by the number of cells in a crossbar row. Reported endurance for RRAM~\cite{Zahoor2020} shows $10^{12}$ writes per cell, which is sufficient for all solutions for ten years. Comparing one-xb and PIMDB, using the aggregation circuit in one-xb does not always improve endurance. This is because one-xb might perform PIM aggregation, whereas PIMDB does pure host aggregation. On Q2.3 and Q4.1, where they both perform PIM aggregation, PIMDB latency is longer and takes more time to carry out the operations. One-xb can achieve the same effect by stalling.
On Q1.1--1.3 and Q3.4, where there are few PIM aggregations for both one-xb and PIMDB, making the latency similar, the lifetime for one-xb is \pimdbEnduGeomean~better in geo-mean. 
The number of writes per cell is even lower for \textit{two-xb} since the PIM operations are distributed across two pages.

\vspace{-4pt}
\section{Related Works}
\vspace{-4pt}
Previous works on bulk-bitwise PIM~\cite{PIMDB,AMBIT,SIMDRAM,CONCEPT,Pinatubo} suggested database applications, and specifically OLAP applications, for bulk-bitwise PIM. These works, however, did not show how a full database benchmark can be performed since they focus on architecture and hardware rather than algorithms. Specifically, these works neither showed how GROUP-BY operations can be performed nor how to address JOIN operations.

PushdownDB~\cite{PushdownDB} showed how to perform GROUP-BY using the available in-cloud database primitives. These primitives are similar to bulk-bitwise PIM primitive database operations: filter and aggregate. Hence, techniques can be borrowed between the domains.

\vspace{-2pt}
\section{Conclusion}
\vspace{-2pt}
This paper presented, for the first time, how bulk-bitwise PIM can perform a full database benchmark, focusing on OLAP database queries. We showed how to adapt a GROUP-BY operation to bulk-bitwise PIM by adding an aggregation circuit and modeling the latency of bulk-bitwise PIM operations. 
We also argue that using pre-joined relations in bulk-bitwise PIM is efficient for OLAP applications. Bulk-bitwise PIM can substantially accelerate execution for single relations, hence storing common pre-joined relations can accelerate the common cases.

Using GROUP-BY and pre-joined relations, we evaluated the performance of bulk-bitwise PIM on the SSB benchmark and compared it to MonetDB, a modern in-memory database. Bulk-bitwise PIM demonstrates a geo-mean speedup of \mntRegSpeedup~and \mntJoinSpeedup, respectively, over the standard and pre-joined MonetDB versions of SSB.

\vspace{-4pt}

\bibliographystyle{IEEEtran}
\bibliography{IEEEabrv,ref}

\begin{thebibliography}{10}
\providecommand{\url}[1]{#1}
\csname url@samestyle\endcsname
\providecommand{\newblock}{\relax}
\providecommand{\bibinfo}[2]{#2}
\providecommand{\BIBentrySTDinterwordspacing}{\spaceskip=0pt\relax}
\providecommand{\BIBentryALTinterwordstretchfactor}{4}
\providecommand{\BIBentryALTinterwordspacing}{\spaceskip=\fontdimen2\font plus
\BIBentryALTinterwordstretchfactor\fontdimen3\font minus
  \fontdimen4\font\relax}
\providecommand{\BIBforeignlanguage}[2]{{%
\expandafter\ifx\csname l@#1\endcsname\relax
\typeout{** WARNING: IEEEtran.bst: No hyphenation pattern has been}%
\typeout{** loaded for the language `#1'. Using the pattern for}%
\typeout{** the default language instead.}%
\else
\language=\csname l@#1\endcsname
\fi
#2}}
\providecommand{\BIBdecl}{\relax}
\BIBdecl

\bibitem{PIMDB}
\BIBentryALTinterwordspacing
B.~Perach \emph{et~al.}, ``{PIMDB: Understanding Bulk-Bitwise Processing
  In-Memory Through Database Analytics},'' 2022. [Online]. Available:
  \url{https://arxiv.org/abs/2203.10486}
\BIBentrySTDinterwordspacing

\bibitem{AMBIT}
V.~Seshadri \emph{et~al.}, ``{AMBIT: In-Memory Accelerator for Bulk Bitwise
  Operations Using Commodity DRAM Technology},'' in \emph{MICRO-50}, 2017.

\bibitem{RACER}
M.~S.~Q. Truong \emph{et~al.}, ``{RACER: Bit-Pipelined Processing Using
  Resistive Memory},'' in \emph{MICRO-54}, 2021.

\bibitem{SIMDRAM}
N.~Hajinazar \emph{et~al.}, ``{SIMDRAM: A Framework for Bit-Serial SIMD
  Processing Using DRAM},'' in \emph{ASPLOS-26}, 2021.

\bibitem{CONCEPT}
N.~Talati \emph{et~al.}, ``{CONCEPT: A Column-Oriented Memory Controller for
  Efficient Memory and PIM Operations in RRAM},'' \emph{IEEE Micro}, vol.~39,
  no.~1, pp. 33--43, 2019.

\bibitem{Pinatubo}
S.~Li \emph{et~al.}, ``{Pinatubo: A Processing-in-Memory Architecture for Bulk
  Bitwise Operations in Emerging Non-Volatile Memories},'' in \emph{DAC-53},
  2016.

\bibitem{Kimball2013}
R.~Kimball and M.~Ross, \emph{The Data Warehouse Toolkit: The Definitive Guide
  to Dimensional Modeling}, 3rd~ed.\hskip 1em plus 0.5em minus 0.4em\relax
  Wiley Publishing, 2013.

\bibitem{Shin2006}
S.~K. Shin and G.~L. Sanders, ``{Denormalization Strategies for Data Retrieval
  from Data Warehouses},'' \emph{Decis. Support Syst.}, vol.~42, no.~1, p.
  267–282, October 2006.

\bibitem{Chirkova2012}
R.~Chirkova and J.~Yang, ``{Materialized Views},'' \emph{Foundations and
  Trends® in Databases}, vol.~4, no.~4, pp. 295--405, 2012.

\bibitem{PushdownDB}
X.~Yu \emph{et~al.}, ``{PushdownDB: Accelerating a DBMS Using S3
  Computation},'' in \emph{ICDE-36}, 2020.

\bibitem{gem5}
N.~Binkert \emph{et~al.}, ``The gem5 simulator,'' \emph{SIGARCH Comput. Archit.
  News}, vol.~39, no.~2, p. 1–7, August 2011.

\bibitem{SSB}
P.~O'Neil \emph{et~al.}, \emph{The Star Schema Benchmark and Augmented Fact
  Table Indexing}.\hskip 1em plus 0.5em minus 0.4em\relax Berlin, Heidelberg:
  Springer-Verlag, 2009, p. 237–252.

\bibitem{MonetDB}
S.~Idreos \emph{et~al.}, ``{MonetDB: Two Decades of Research in Column-oriented
  Database Architectures},'' \emph{{IEEE} Data Eng. Bull.}, vol.~35, no.~1, pp.
  40--45, 2012.

\bibitem{Dreseler2020}
M.~Dreseler \emph{et~al.}, ``Quantifying tpc-h choke points and their
  optimizations,'' \emph{Proc. VLDB Endow.}, vol.~13, no.~8, p. 1206–1220,
  April 2020.

\bibitem{Rabl2013}
T.~Rabl \emph{et~al.}, ``{Variations of the Star Schema Benchmark to Test the
  Effects of Data Skew on Query Performance},'' in \emph{ICPE-4}, 2013.

\bibitem{Xu2015}
C.~{Xu} \emph{et~al.}, ``{Overcoming the challenges of crossbar resistive
  memory architectures},'' in \emph{HPCA-21}, 2015.

\bibitem{Dageville2004}
B.~Dageville \emph{et~al.}, ``{Automatic SQL Tuning in Oracle 10g},'' in
  \emph{VLDB-30}.\hskip 1em plus 0.5em minus 0.4em\relax VLDB Endowment, 2004.

\bibitem{PIMConsistency}
B.~Perach \emph{et~al.}, ``On consistency for bulk-bitwise
  processing-in-memory,'' in \emph{HPCA-29}, 2023.

\bibitem{simulation}
\BIBentryALTinterwordspacing
B.~Perach, ``gem5 bulk-bitwise {PIM} consistency.'' [Online]. Available:
  \url{https://github.com/benperach/gem5_bulkbitwise_PIM_consistency}
\BIBentrySTDinterwordspacing

\bibitem{Talati2016}
N.~Talati \emph{et~al.}, ``{Logic Design Within Memristive Memories Using
  Memristor-Aided loGIC (MAGIC)},'' \emph{IEEE Tran. on Nanotechnology},
  vol.~15, no.~4, pp. 635--650, 2016.

\bibitem{NVSim}
X.~{Dong} \emph{et~al.}, ``{NVSim: A Circuit-Level Performance, Energy, and
  Area Model for Emerging Nonvolatile Memory},'' \emph{IEEE Tran. on
  Computer-Aided Design of Integrated Circuits and Systems}, vol.~31, no.~7,
  pp. 994--1007, 2012.

\bibitem{Zahoor2020}
F.~Zahoor \emph{et~al.}, ``{Resistive Random Access Memory (RRAM): an Overview
  of Materials, Switching Mechanism, Performance, Multilevel Cell (mlc)
  Storage, Modeling, and Applications},'' \emph{Nanoscale Research Letters},
  vol.~15, April 2020.

\end{thebibliography}

\end{document}